\documentclass{aastex}
\usepackage{spr-astr-addons}
\usepackage{url}

\RequirePackage{color}

\begin{document}

\title{Thin accretion disc with a corona in a central magnetic field}
\shorttitle{Disc with a corona}
\shortauthors{F. Khajenabi, M. Shadmehri, S. Dib}

\author{Fazeleh Khajenabi\altaffilmark{1}} \and \author{Mohsen Shadmehri\altaffilmark{2,3}}
\affil{fazeleh.khajenabi@ucd.ie, mohsen.shadmehri@dcu.ie}
\and
\author{Sami Dib\altaffilmark{4,5}}
\affil{sami.dib@cea.fr}

\altaffiltext{1}{School of Mathematical Sciences, University College Dublin, Belfield, Dublin 4, Ireland}
\altaffiltext{2}{School of Mathematical Sciences, Dublin City University, Glasnevin, Dublin 9, Ireland}
\altaffiltext{3}{Department of Physics, School of Science, Ferdowsi University, Mashhad, Iran}
\altaffiltext{4}{Lebanese University, Faculty of Sciences, Department of Physics, El Hadath, Beirut, Lebanon }
\altaffiltext{5}{Service d'Astrophysique, DSM/IRFU/SAp, CEA/Saclay, 91191, Gif-sur-Yvette Cedex, France}

\begin{abstract}
We study the steady-state structure of an accretion disc with a
corona surrounding a central, rotating, magnetized star. We assume
that the magneto-rotational instability  is the dominant mechanism
of angular momentum transport inside the disc and is responsible for
 producing magnetic tubes above the disc. In our model, a  fraction of the
dissipated energy inside the disc is transported to the corona via
these magnetic tubes. This energy exchange from the disc to the
corona which depends on the disc physical properties is modified
because of the magnetic interaction between the stellar magnetic
field and the accretion disc.  According to our fully analytical
solutions for such a system, the  existence  of a corona not only
increases the surface density but reduces the temperature of the
accretion disc. Also, the presence of a corona enhances the ratio of
gas pressure to the total pressure. Our solutions show that when the strength of
the magnetic field of the central neutron star is large or the star is rotating fast enough,  profiles of the physical variables of the disc significantly modify due to the existence of a corona.

\end{abstract}


\section{Introduction}
Accretion discs are believed to be present in a wide variety of
astronomical systems.  The standard theory of geometrically  thin
accretion discs has provided successful framework for understanding
the basic physical properties of these objects (Shakura \& Sunyaev
1973). A solution for the structure of an accretion disc in an
external magnetic field is desirable because accretion discs are
frequently found around magnetic stars (Verbunt 1993; Warner 1995;
Romanova et al 2003). The stellar magnetic fields interacts with the
disc and can allow the transfer of angular momentum between the disc
and the star and this significantly affects the structure of the
disc and the spin evolution of the star (e.g. Ghosh \& Lamp 1978;
Aly 1980; Wang 1987; Armitage, Clarke \& Tout 1999; Schenker et al
2002; Dai \& Li 2006). Matthews et al (2005) reformulated the
standard disc solution to incorporate the effect of a torque from a
stellar magnetic field. They studied  the effects of varying  the
mass transfer rate, spin period and strength of the stellar magnetic
field using their analytical solutions. Also, there are several
simulations of the interactions between the stellar magnetic filed
and accretion disc (e.g. Miller \& Stone 1997; Romanova et al 2002;
Romanova et al 2003). Miller \& Stone (1997) investigated disc-star
interaction for different geometries and stellar magnetic fields.
Romanova et al (2002) described in detail the disc-star interaction
for the case of slowly rotating stars and also cases of fast
rotating stars.

On the other hand, there are varying models of a hot corona above a
cool disc  (e.g. Haardt\& Maraschi 1993). Begelman, McKee \& Shields   (1983)
presented detailed  calculations of the formation of the corona by
irradiation from the central source. Rozanska \& Czerny (1996) and
Rozanska et al (1999) investigated the effect of illumination by the
central source on the accretion disc for either a black hole as in
AGN or a neutron star as in X-ray binary sources. It was found that
the disc and the corona are separated by a boundary layer or heated
upper region of the disc, i.e. the region above the disc becomes
stratified with increasing temperature and decreasing density (see
also Rozanska et al 2002).  Jimenez-Garate, Raymond \& Liedahi (2002) calculated
the vertical structure  of an Accretion Disc Corona (ADC) and
allowed the radius of the corona to extend to the same radius of the
accretion disc. Low Mass X-ray Binaries (LMXBs) have  two continuum
components of radiation: simple blackbody  emission from the neutron
star plus comptonized emission from an extended  corona above
accretion disc (e.g. Church \& Balucinska-Church 1995). Church \&
Balucinska (2004) have demonstrated the importance of the size of
the ADC to the correct description of Comptonization and they
derived the comptonized spectrum of an LMXB based on the thermal
Comptonization of  seed photons. They estimated the radial extension
of ADC in some LMXBs up to $10^{9} \rm cm$.

Generally, models for generating an ADC can be divided into two
groups:  (1) The dissipation of acoustic or magnetic energy flux
generated by convective or viscous turbulence within the disc (e.g. Galeev, Rosner \& Vaiana 1979; Merloni \& Fabian 2002; Merloni  2003); or
(2) the evaporation of disc material by the central source (e.g.
Begelman, McKee \& Shields 1983).

Differential rotation of a standard accretion disc amplifies the
magnetic field.  This may result in a magnetic active corona
surrounding the disc (Burm 1986). These magnetic loops which thread
the disc are analogous to the solar case. The problem of magnetic
flux escape from an accretion disc has been studied by many authors
(Stella \& Rosner 1984; Heyvaerts \& Priest 1989; Shibata, Tajima \&
Matsumoto 1990; Romanova et al 1998).

The are reasons to believe that some parts of the ADC may find
itself embedded in the field of the  central object. Stollman \&
Kuperus (1988) assumed that the neutron stars in some of the LMXB
have magnetic fields, which are capable of holding the accretion
disc at some distance from the stars surface. They developed a model
for the interaction of the magnetic field of the neutron star and the
magnetic loops in the corona above the disc. Flaring interactions
between accretion disc and neutron star magnetosphere has also been
studied by many authors (e.g. Aly \& Kuijpers 1990). Magnetic loops of
the accretion disc which extend toward corona may contribute to
angular momentum transport in the disc (Burm \& Kuperus 1988).
However, the possible effects of the corona on the structure of a
thin accretion disc under the influence of a stellar magnetic field
have not been studied to our knowledge. In this paper, we extended
Matthews et al (2005) analysis by considering the corona of the disc
as well. The following question is addressed: what is the effect of
the corona on the steady state structure of a magnetically torqued
thin accretion disc?  We  obtain a set of analytical solutions for a
thin accretion disc with a corona, to which a magnetic torque due to
the central object (e.g. neutron star) is applied. We perform a
parameter study to explore how physical variables of the disc such
as density or temperature may be modified due to  changes in stellar
magnetic field or stellar spin. In section 2, we discuss the  basic
assumptions and present the main equations of the model. Typical
physical properties of the solutions will be studied in section 3.
In the final section we conclude the paper by a summary of our results.

\section{General formulation}
We assume that all angular momentum transport takes place in the disc and
the mass accretion rate $\dot{M}$ is constant with radius and time. However, the microphysics mechanism of the angular
 momentum transfer remains unknown.  Shakura \& Sunyaev (1973) replaced all the missing physics by a parameter $\alpha$.
 This approach has been widely used for studying the dynamics and structure of  accretion flow. A promising mechanism for
 driving the turbulence responsible for angular momentum and energy transport is the action of the magneto-rotational
 instability (MRI) that is expected to take place in such discs (Balbus \& Hawley 1991). But for a disc-corona
 system,
 it is typically thought that the viscous stress, assumed to be magnetic in nature, transports angular momentum and
 initially randomise the gravitational binding energy near the midplane. The magnetized fluid elements, which are
 buoyant with respect to their surroundings, dissipate above the disc. Merloni \& Fabian (2002) and Merloni (2003)
 presented  very detailed discussions about dissipation in the corona and its relation with angular momentum transport
 in the disc itself. Our disc-corona model is developed along the line proposed by Merloni \& Fabian (2002) and Merloni (2003).

We  consider a more general prescription for the viscous stresses $\tau_{\rm r\phi}$ (Taam \& Lin 1984; Watarai \& Mineshige 2003; Merloni \&  Nayakshin 2006):
\begin{equation}
\tau_{\rm r\phi}=-\alpha_{0} p^{1-\mu/2} p_{\rm gas}^{\mu/2},\label{eq:visg}
\end{equation}
where $\alpha_{0}$ and $0\leq\mu\leq2$ are constants. Also, $p$ is
the sum of the gas and radiation pressures. Phenomenological  models
generally assume that at each radius, a fraction $f$ of accretion
energy is released in the reconnected magnetic corona. Assuming that
turbulence inside the disc is MRI-driven, such a fraction $f$ of the
binding energy is transported from large to small depths by Poynting
flux, Merloni \& Nayakshin (2006) estimated the fraction $f$ as
\begin{equation}
f=\sqrt{2\alpha_{0} \beta^{\mu/2}},\label{eq:f}
\end{equation}
where $\beta$ is the ratio of gas pressure to the total pressure. So, in this model, this fraction $f$ is not a free parameter. Actually, it is very difficult to present a detailed analytical  model which gives us the fraction $f$ as a function of the disc physical variables structure and MRI (even without central magnetic torque). We appreciate on possible effect of central magnetic torque on energy exchange $f$. But it is beyond our study because of severe limitations of our phenomenological approach. But we would like to emphasize that the amount of the energy exchange  corresponding to  solutions with central magnetic torque (our solutions) and without central torque is not the same. Because for solutions with a central magnetic torque, we have an extra magnetic torque in angular momentum equation (see equation (\ref{eq:PhidirecA}))which dramatically changes the solutions comparing to the solutions without that term (see solutions of Matthews et al. (2005)). In fact, equation (\ref{eq:f}) prescription is based on MRI inside of the disc, irrespective of the (non)existence of external magnetic torque. So, with central magnetic torque its effect on the energy exchange appears by changing disc structure (through an extra magnetic term in angular momentum equation).

The rotation curve is dominated by a Newtonian point mass $M$, as
relativistic effects are only important at small radii. Thus, the rotational angular velocity of the disc is
Keplerian, i.e. $\Omega_{\rm K}=\sqrt{GM/R^{3}}$. Having equation
(\ref{eq:visg}) as a prescription for the viscous stresses and
equation (\ref{eq:f}) as an expression for the fraction of power
dissipated in the corona, we can write the basic equations of the
disc. Vertical hydrostatic equilibrium of the disc implies
\begin{equation}\label{eq: Zdirec}
\frac{p}{\Sigma}=\frac{\Omega_{\rm K}^{2}H}{2}.
\end{equation}

\begin{figure}
\plotone{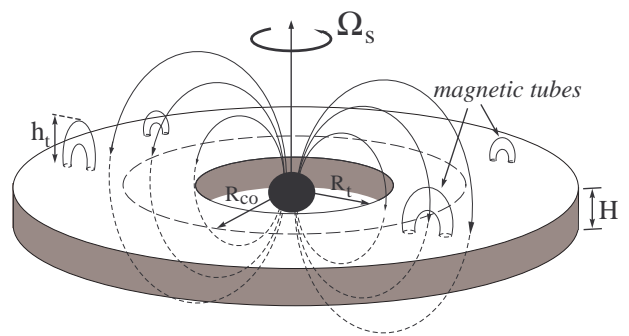}
\caption{Schematic of the
magnetic field of
 the neutron star and magnetic tubes of the accretion disc. The truncation and the corotation
 radii are marked by $R_{\rm t}$ and $R_{\rm co}$, respectively. }\label{fig:fig}
\end{figure}

In order to write the azimuthal component of equation of motion, the specific magnetic torque of the central neutron star can be written as
\begin{equation}
\Lambda  = \frac{\sqrt{GM}R^{1/2}}{t_{\Lambda}},\label{eq:torque}
\end{equation}
where $t_{\Lambda}$ is the time-scale on which the local disc
material gains angular momentum. This torque appears because of the
interaction of the field lines of the central, rotating magnetized
neutron star  with its surrounding partially ionized inner accretion
disc. Actually neither the physics of angular momentum transport by
the magnetic field of the central star, nor the mechanism of
penetration of the magnetic field into the accretion disc are well
understood. Nevertheless we can present some physical prescriptions
for the magnetic torque of the central neutron star. Actually, there
are some interactions between the magnetic tubes above the disc and
the magnetic field of the neutron star. However, we concentrate on
the steady structure of the disc and  such complicated interactions
are neglected in our model. Formation of the corona is a direct
consequence of the magnetic tubes of the accretion disc. On the
other hand, the physical properties of the accretion disc are
modified because of the existence of the magnetic torque and  the
dissipated energy into the corona depends on the structure of the
disc. Thus, the effect of the stellar magnetic field on the magnetic
tubes is included via its torque on the disc indirectly in our
model. Figure \ref{fig:fig} show schematically the magnetic field of
the neutron star and the magnetic tubes of the disc. The magnetic
structures in the corona of the disc should be of a small scale,
because turbulent eddies in the disc are believed to have the disc
thickness as a characteristic size, as should the magnetic loops
which pop out as a result of magnetic buoyancy. However, there is a
possibility of forming larger magnetic loops above the disc by
reconnecting small scale loops (e.g., Heyvaerts \& Priest 1984). Analysis of such complexities is
beyond this paper.

In equation (\ref{eq:torque}), the torque time-scale $t_{\Lambda}$
is equivalent to the magnetic time-scale $t_{\rm mag}$. Magnetic
field of the neutron star has a dipolar form, i.e. $B_{\rm z} \sim
|{\bf \mu}| R^{-3}$ where ${\bf \mu}$ is the magnetic moment of the
star. Depending on the magnetic interaction model, the magnetic
time-scale can be parametrized in the form (Livio \& Pringle 1992;
Campbell \& Heptinstall 1998; Matthews et al 2005)
\begin{equation}
t_{\rm mag}= \frac{2\Sigma}{\beta_{M}}\frac{R^{\gamma
+2}}{[(R/R_{co})^{3/2} -1]},
\end{equation}
where $\gamma$ is determined by the magnetic interaction model
(e.g., Campbell \& Heptinstall 1998) and $\beta_{M}$ is calculated
by the mass and the magnetic moment of the star,
\begin{equation}
\beta_{M}=\frac{|{\bf \mu}|^2}{2\pi\sqrt{GM}}.
\end{equation}
In our fully magnetized case we have $\gamma=7/2$ (Livio \& Pringle
1992; Matthews et al 2005). Also, in the above equation, the
corotation radius is represented by $R_{co}$ which is by definition
the radius of the accretion disc at which, the star and the disc rotate at the same angular rate (Figure \ref{fig:fig}). Note that the dipole and spin axes are taken to be parallel to $z$, and the whole system is assumed to be axisymmetric.

Now, we can write the azimuthal component of the equation of motion
as
\begin{equation}\label{eq:PhidirecA}
\Sigma v_{\rm R}\frac{d}{dR}(R v_{\rm\varphi})=\frac{1}{R}\frac{d}{dR}(R^{2}\tau_{r\phi})+\Sigma \Lambda,
\end{equation}
where $v_{\rm R}$ and $v_{\varphi}$ are the radial and the rotational velocities of the disc. We note that the disc is rotating with Keplerian profile, i.e. $v_{\rm\varphi}=v_{\rm K}$. The $r-\varphi$ component of viscous stress tensor is prescribed by equation (\ref{eq:visg}). Also, the parameter $\Lambda$ is representing the specific torque of the central neutron star which is given by equation (\ref{eq:torque}). After mathematical manipulations, we can write equation (\ref{eq:PhidirecA}) as
\begin{equation}\label{eq:Phidirec}
8\pi\alpha_{0} H (p_{\rm gas})^{\mu/2} p^{(2-\mu)/2}= 3\Omega_{\rm K} \dot{M} q^{4}v,
\end{equation}
where,
\begin{displaymath}
v=1-\frac{\beta_{M}} {\dot{M}}\frac{\pi
h}{R^{\gamma}(\gamma-2)q^{4}},
\end{displaymath}

\begin{displaymath}
h=(\frac{R}{R_{co}})^{\frac{3}{2}}[1-(\frac{R_{t}}{R})^{2-\gamma}]-\frac{\gamma-2}{\gamma-(\frac{1}{2})}[1-(\frac{R_{t}}{R})^{\frac{1}{2
}-\gamma}],
\end{displaymath}
\begin{displaymath}
q=[1-(\frac{R_{t}}{R})^{\frac{1}{2}}]^{\frac{1}{4}}.
\end{displaymath}
The truncation radius $R_t$ denotes the inner boundary of the
accretion disc. A common feature of all magnetic star-disc
interaction models is that the stellar magnetosphere is strong
enough to disrupt the accretion disc at some radius $R_{t}$ above
the surface of the star.  There are several methods to locate the
truncation radius. We use the approach of Matthews et al (2005).
They presented an approximate relation for $\gamma=7/2$ as
\begin{equation}
\frac{R_t}{R_{co}}\simeq
[\frac{Q}{2}(\sqrt{\frac{4}{Q}+1}-1)]^{2/3},
\end{equation}
where the nondimensional parameter $Q$ is determined by the ratio
$\beta_{M}$, accretion rate and the corotation radius, i.e.
\begin{equation}
Q=2\pi R_{co}^{-\gamma}(\frac{\beta_{M}}{\dot{M}}).
\end{equation}
The truncation and corotation radii are shown schematically in
Figure \ref{fig:fig}.

The energy equations is given by
\begin{equation}\label{eq:energy}
\sigma T_{\rm eff}^{4}=\frac{3}{8\pi}\Omega_{\rm K}^{2}\dot{M} q^{4}v (1-f),
\end{equation}
where as we discussed $f=\sqrt{2\alpha_{0}\beta^{\mu/2}}$. Also, we
can assume that the vertical transport of heat is by radiative
diffusion which implies the midplane and the surface temperatures
are related by

\begin{equation}
T = (\frac{3}{8}\kappa\Sigma)^{1/4} T_{\rm eff},
\end{equation}
where $\kappa$ is the opacity coefficient. In order to limit the
parameter space in our analysis, we simply assume the electron
opacity for this coefficient, i.e. $\kappa = \kappa_{e}= 0.4$ cm$^2$ g$^{-1}$.

Now, equations (\ref{eq: Zdirec}), (\ref{eq:Phidirec}) and
(\ref{eq:energy}) are the main equations which enable us to find $p$
and $T$ as functions of $R$ and $\beta$ and the other input
parameters. Thus,
\begin{displaymath}
T=(\frac{4\sigma \Omega_{\rm K}}{3\kappa \alpha_{0}})^{-1/2}(\frac{16\pi^{2}\alpha_{0}^{2}ck_{\rm B}}{3\sigma \mu_{\rm m} m_{\rm H} \dot{M}^{2}\Omega_{\rm K}^{4} q^8 v^2})^{-1/3}
\end{displaymath}
\begin{equation}\label{eq:main1}
\times \frac{(1-\sqrt{2\alpha_{0}\beta^{\mu/2}})^{1/2}}{
(1-\beta)^{1/3}}\beta^{(4-\mu)/12},
\end{equation}
\begin{displaymath}
p=(\frac{4\sigma \Omega_{\rm K}}{3\kappa \alpha_{0}})^{-1/2}(\frac{16\pi^{2}\alpha_{0}^{2}ck_{\rm B}}{3\sigma \mu_{\rm m} m_{\rm H} \dot{M}^{2}\Omega_{\rm K}^{4} q^8 v^2})^{-2/3}
\end{displaymath}
\begin{equation}\label{eq:main2}
\times\frac{(1-\sqrt{2\alpha_{0}\beta^{\mu/2}})^{1/2}}{
(1-\beta)^{2/3}}\beta^{(8-5\mu)/12}.
\end{equation}
There is an algebraic equation for $\beta$ as follows
\begin{displaymath}
\frac{k_{\rm B}}{\mu_{\rm m} m_{\rm H}} (\frac{4\sigma\Omega_{\rm K}}{3\kappa \alpha_{0}})^{-3/2}
(\frac{16\pi^{2}\alpha_{0}^{2}c k_{\rm B}}{3\sigma\mu_{\rm m} m_{\rm H} \dot{M}^{2}\Omega_{\rm K}^{4} q^{8} v^{2}})^{-5/3}
\end{displaymath}
\begin{equation}\label{eq:beta}
 \times (\frac{8\pi \alpha_{0}}{3\Omega_{\rm K}^{2}\dot{M} q^{4} v})^{2} \frac{(1-\sqrt{2\alpha_{0}\beta^{\mu/2}})^{3/2}}{(1-\beta)^{5/3}}\beta^{(8+\mu)/12}=1,
\end{equation}
where $\mu_{\rm m}$ is the mean particle mass in units of the hydrogen atom mass, $m_{\rm H}$. The other constants have their usual meanings.

 \begin{figure*}
\vspace*{+100pt}
\includegraphics[scale=0.8]{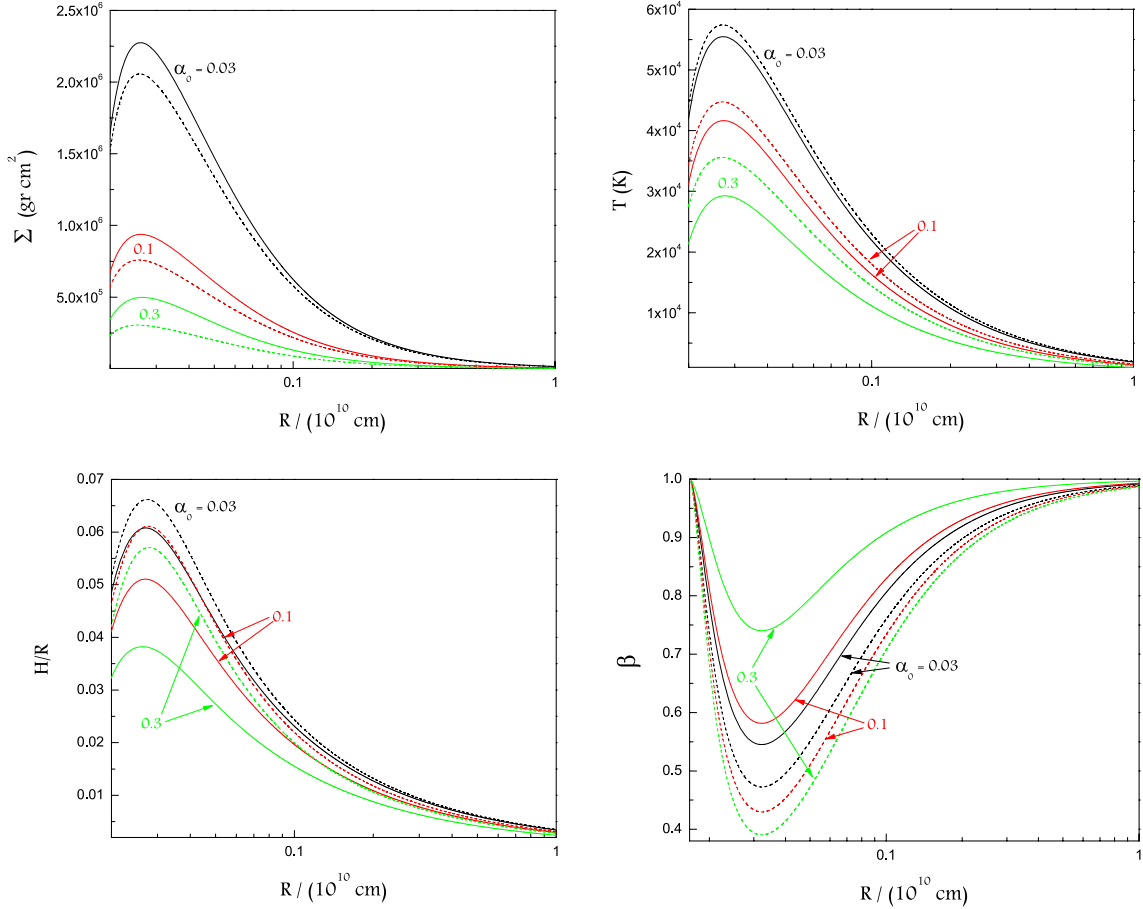} \caption{Profiles of the
physical variables  for a disc with a corona ({\it solid lines}) and
without a corona ({\it dashed lines})  vs. radial location in the
disc with $\mu=1.0$, $\mu_{\rm m}=0.6$, $\hat{k}=1$, $M_{1}=1.4$,
$\dot{M}_{16}=1$, $P_{\rm spin} = 1$ s and $B_{s}=10^{12}$ G. Each
curve is labeled by its corresponding viscosity coefficient
$\alpha_0$. Also, in the online version of the paper, curves are
shown by different colors depending on the viscosity coefficient.
The top left-hand and right-hand plots show the surface density and
the temperature of the accretion disc, respectively. The ratio of
the thickness of the disc and the radius is shown in the bottom
left-hand plot. Finally, the bottom right-hand plot shows profile of
the ratio $\beta$ of gas pressure to total pressure. Generally, we see that the corona becomes more
effective when the viscosity coefficient $\alpha_0$ has larger
values.}\label{fig:figure2}
\end{figure*}
\begin{figure*}
\vspace*{+100pt}
\includegraphics[scale=0.8]{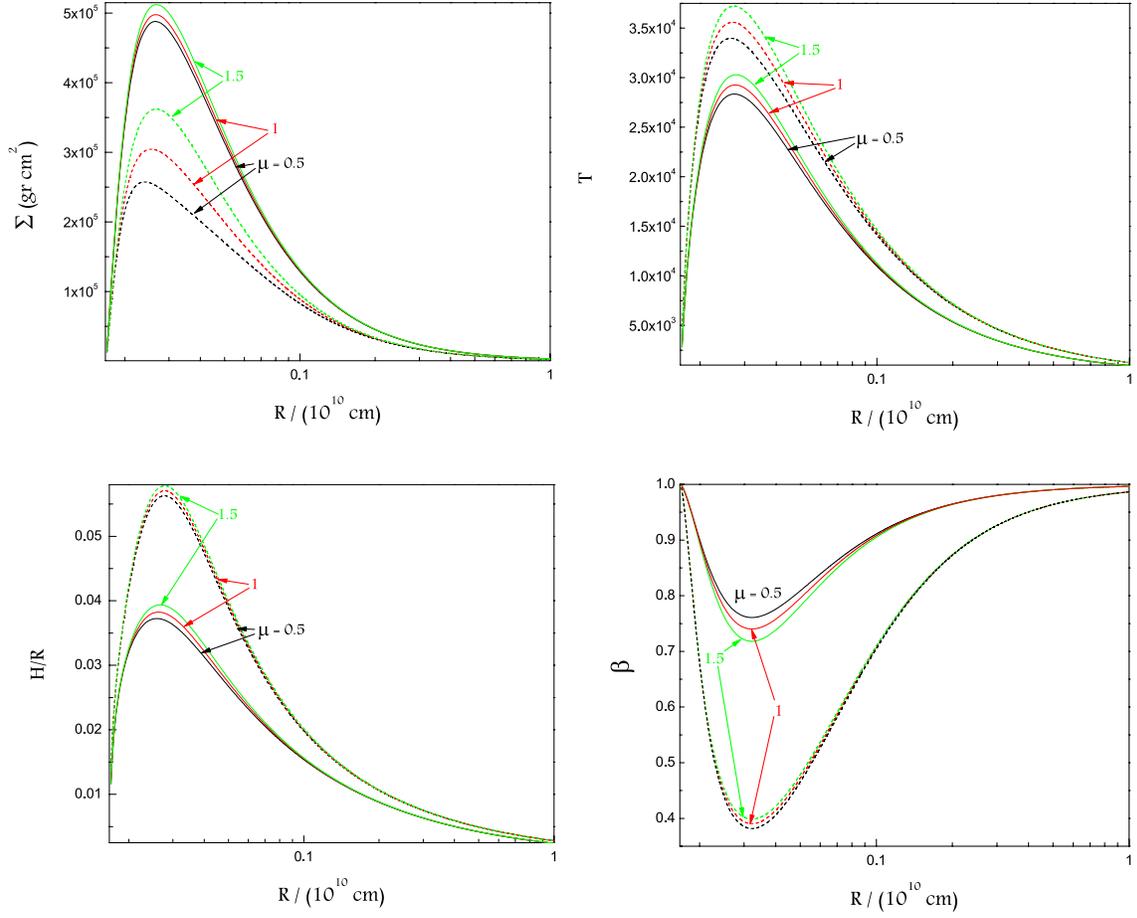}\caption{Profiles of the
physical variables  for a disc without a  corona ({\it dashed
lines}) and with a corona ({\it solid lines})  vs. radial location
in the disc  with $\alpha_0=0.3$, $\mu_{\rm m}=0.6$, $\hat{k}=1$,
$M_{1}=1.4$, $\dot{M}_{16}=1$, $P_{\rm spin} = 1$ s and
$B_{s}=10^{12}$ G. These plots show dependence of the physical
variables to the viscosity exponent $\mu$. Typical behaviors of the ratio $H/R$ and the ratio  of the gas pressure to the total pressure, $\beta$,
for the case without a corona are not very sensitive to the exact value
of the viscosity exponent. But in the case of nonexistence of a
corona, profiles of the surface density and the temperature of the
accretion disc are not independent of the parameter $\mu$. As this
exponent increases, both the surface density and the temperature
increase.}\label{fig:figure3}
\end{figure*}

\begin{figure*}
\vspace*{+100pt}
\includegraphics[scale=0.8]{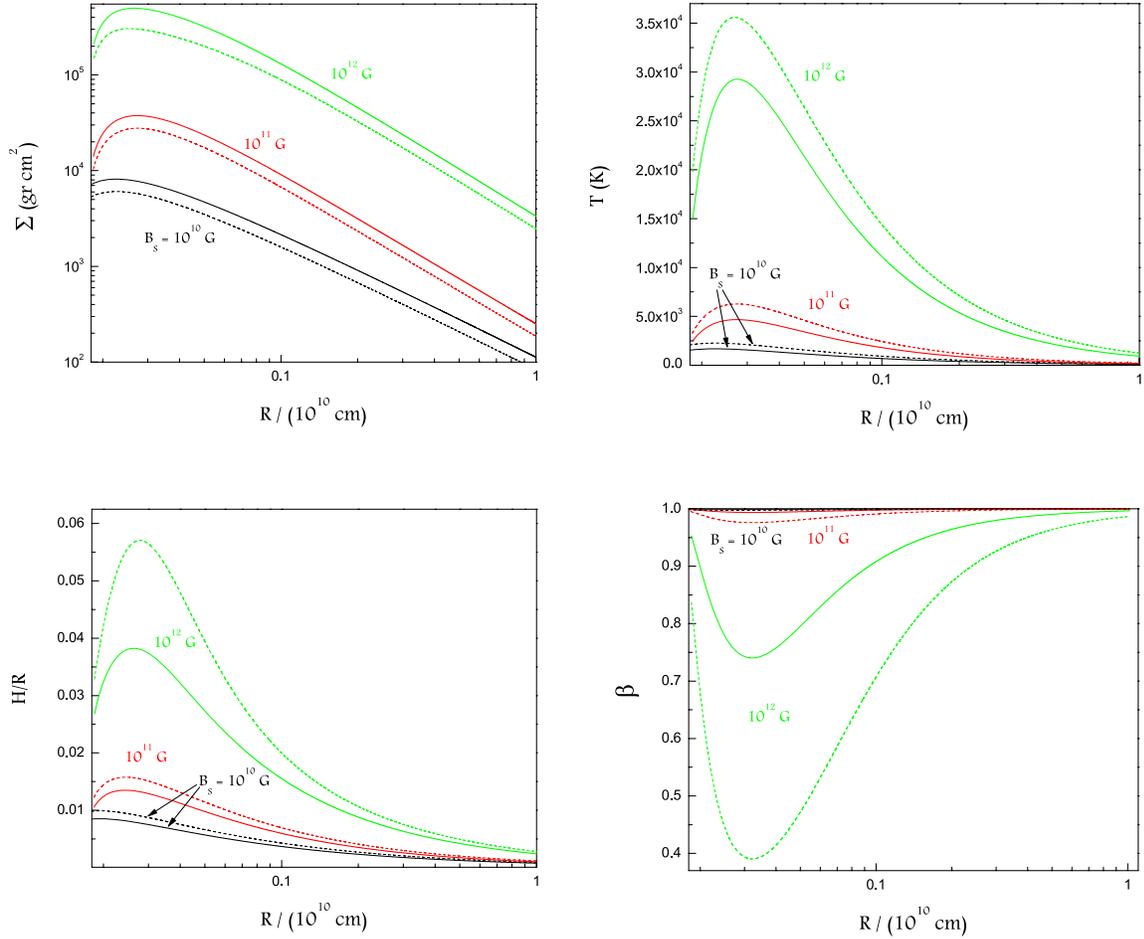} \caption{Profiles of the
physical variables  for a  disc without a corona ({\it dashed
lines}) and with a corona ({\it solid lines})  vs. radial location
in the disc  with $\alpha_0=0.3$, $\mu=1.0$, $\mu_{\rm m}=0.6$,
$\hat{k}=1$, $M_{1}=1.4$, $\dot{M}_{16}=1$ and $P_{\rm spin} = 1$ s.
Effects of the magnetic field of the central neutron star are shown
in these plots. Each curve is labeled by the magnetic field at the
surface of the neutron star, i.e. $B_{s}$. As the magnetic field
increases, the possible differences between the solutions with a
corona and without a corona become more
significant. The bottom right-hand plot shows profile of
the ratio $\beta$ of gas pressure to total pressure.}\label{fig:figure4}
\end{figure*}
\begin{figure*}
\vspace*{+100pt}
\includegraphics[scale=0.8]{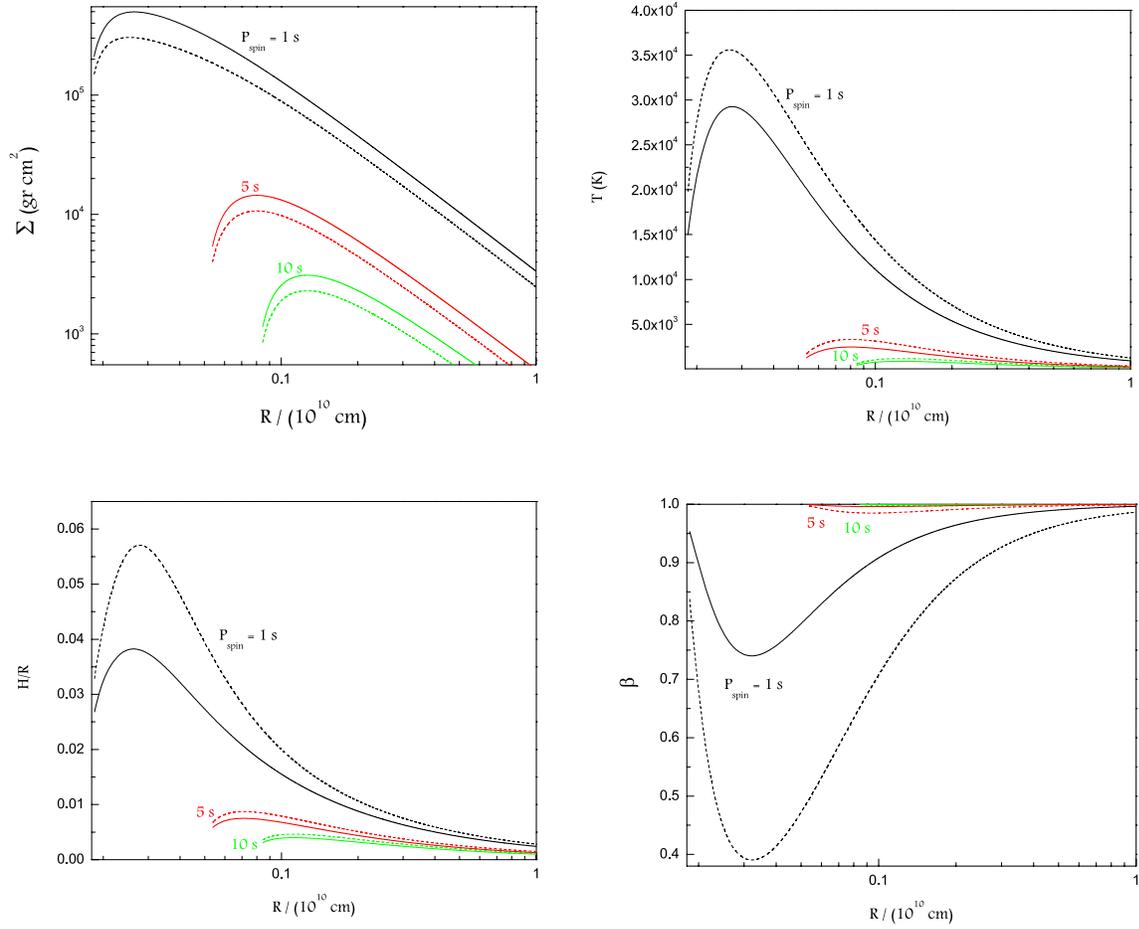}
\caption{Profiles of the physical variables  for a disc  without a
corona ({\it dashed lines}) and with a corona ({\it solid lines})
vs. radial location in the disc  with $\alpha_0=0.3$, $\mu=1.0$,
$\mu_{\rm m}=0.6$, $\hat{k}=1$, $M_{1}=1.4$, $\dot{M}_{16}=1$ and $B_{s}=10^{12}$ G. The bottom right-hand plot shows profile of
the ratio $\beta$ of gas pressure to total pressure.
The central magnetic neutron star is rotating at different
speed and each curve is labeled by the rotational period of the
star. We see that the truncation radius increases as the central
neutron star is rotating slower. }\label{fig:figure5}
\end{figure*}

In order to study the behavior of our solutions, it is more
convenient to introduce dimensionless variables. For the central
mass $M$,  we introduce $M_1=M/M_{\odot}$ and for opacity we assume $\hat{\kappa}=\kappa / \kappa_{e}$. The radial distance from
the star, in the plane of the disc, is represented by $R$ in units
of  $10^{10}$ cm. So, we introduce $r_{10}=R/(10^{10} \rm cm)$.  Also $\dot{M}_{16}=\dot{M}/(10^{16} {\rm g
s}^{-1})$ represents the mass accretion rate in units of  $10^{16}$
g s$^{^{-1}}$ .

We can rewrite our solutions as
\begin{displaymath}
 \rho=2.63 \times10^{-10} \alpha_{0}^{-1/2}  \hat{\kappa}^{3/2} M_{1}^{5/4} \dot{M}_{16}^{2}   q^{8} v^{2} r_{10}^{-15/4}
\end{displaymath}
\begin{equation}
\times \beta^{(8-\mu)/4}(1-\beta)^{-2}(1-\sqrt{2\alpha_{0}\beta^{\mu/2}})^{3/2},\label{eq:rho}
\end{equation}

\begin{displaymath}
p=40.47\times\alpha_{0}^{-5/6}\hat{\kappa}^{1/2} M_{1}^{13/12} \dot{M}_{16}^{4/3}  q^{16/3} v^{4/3} r_{10}^{-13/4}
\end{displaymath}
\begin{equation}
 \times \beta^{(8-5\mu)/12}(1-\beta)^{-2/3}(1-\sqrt{2\alpha_{0}\beta^{\mu/2}})^{1/2},\label{eq:p}
\end{equation}

\begin{displaymath}
 \frac{H}{R}=0.24\times \alpha_{0}^{-1/6}\hat{\kappa}^{-1/2} M_{1}^{1/6} \dot{M}_{16}^{-1/3} q^{-4/3} v^{-1/3}
\end{displaymath}
\begin{equation}
 \times \beta^{-(8+\mu)/12}(1-\beta)^{2/3}(1-\sqrt{2\alpha_{0}\beta^{\mu/2}})^{-1/2},\label{eq:HR1}
\end{equation}
and the ratio $\beta$ is obtained from the non-dimensional form of equation (\ref{eq:beta}), that is,

\begin{displaymath}
 4.69 \times 10^{-7} \alpha_{0}^{1/6} \hat{\kappa}^{3/2} M_{1}^{7/12} \dot{M}_{16}^{4/3} q^{16/3} v^{4/3} r_{10}^{-7/4}
\end{displaymath}
\begin{equation}
 \times \beta^{(8+\mu)/12}(1-\beta)^{5/3}(1-\sqrt{2\alpha_{0}\beta^{\mu/2}})^{3/2}= 1.\label{eq:HR2}
\end{equation}
We can also calculate the surface density as

\begin{displaymath}
 \Sigma=2.12 \times 10^{-4} \alpha_{0}^{-2/3} \hat{\kappa} M_{1}^{17/12} \dot{M}_{16}^{5/3} q^{20/3} v^{5/3} r_{10}^{-11/4}
\end{displaymath}
\begin{equation}
 \times \beta^{(4-\mu)/3}(1-\beta)^{-4/3}(1-\sqrt{2\alpha_{0}\beta^{\mu/2}})^{1/2}.\label{eq:HR3}
\end{equation}

Typical disc structure will now be illustrated, using the full disc solutions quoted in equations (\ref{eq:rho}), (\ref{eq:p}) and (\ref{eq:HR1}). However, the physical variables depend not only on the radial distance but also on the ratio of the gas pressure to the total pressure which can be calculated at each radius from algebraic equation (\ref{eq:HR2}). Having analytical solutions, we can study physical properties of our disc and corona system. In the next section, we will do a parameter study of our solutions.

\section{Analysis}

We fix the central mass, the mass transfer rate, the opacity  and
the mean molecular weight, respectively as $M_{1}=1.4$,
$\dot{M}_{16}=1$, $ \hat{\kappa}=1$  and  $\mu_{m}=0.6$. The other
input parameters are changed to illustrate their possible effects on
the physical properties of the system. Typical behaviors of the
solutions for disc with a corona (solid curves) and without a corona
(dashed curves)  are shown in all subsequent  figures. Each curve is
labeled by its corresponding parameter, though color versions are
available in the online version of the paper. Note that the radial
extension of the plots are up to $10^{10}$ cm, significant effects
of the corona and the central magnetic field are seen at radii less
than $10^9$ cm, generally.

Figure $\ref{fig:figure2}$ shows typical behavior of the physical
quantities  of the accretion disc when the viscosity coefficient
$\alpha_0$ changes from low value $0.03$ to high value $0.3$. The
profile of the surface density is shown in the top left-hand plot of
Figure $\ref{fig:figure2}$. When the viscosity coefficient
increases, the surface density decreases both in the cases with and
without a corona. This behavior can be explained simply by the fact the the accretion rate is constant in our model. So, the radial velocity (which is proportion to viscosity coefficient) is inversely proportional to the surface density. Thus, when the viscosity coefficient increases, the radial velocity increases too. But the surface density decreases with the viscosity coefficient. However, the corona significantly  increases the surface density. The top right-hand plot of Figure
$\ref{fig:figure2}$ shows the presence of a corona serves to cool
the disc. In fact,  corona extracts some of the dissipated energy inside the disc. When there is not a corona, all of the dissipated energy heats the disc. But with a corona some of the dissipated energy transfers from the disc to the corona and so, the disc becomes cooler. Generally, when the viscosity coefficient increases the
disc temperature decreases both with and without a corona. In fact,
the fraction of the  dissipated energy into the corona is directly
proportional to $\alpha_0$ according to  equation (\ref{eq:f}).
Thus, as the profile of the temperature shows, the existence of a
corona leads to a cooler disc and the corona becomes more effective
when the viscosity has larger values. The bottom left-hand plot of
Figure $\ref{fig:figure2}$ shows the ratio of the disc thickness to the
radius. The corona causes the disc to become thinner than the case
without a corona. It justifies the thin disc approximation even in
the presence of a corona. The ratio of the disc thickness to the
radius decreases when the viscosity coefficient increases.
Hydrostatic equilibrium of the disc in the vertical direction
implies that the thickness of the disc is in proportion to the sound
speed (i.e. temperature). Since the existence of the corona reduces
the temperature of the disc, larger values of the viscosity
coefficient imply a thinner disc.

The bottom right-hand plot of $\ref{fig:figure2}$ shows the ratio
$\beta$ of  gas pressure to the total pressure of the disc. In the
absence of a corona, when the viscosity coefficient increases, the
ratio of gas pressure to the total pressure  decreases (i.e.,
radiation pressure increases). However, in a disc with a corona, the
parameter $\beta$  increases as the viscosity coefficient increases
(i.e., radiation pressure decreases). Generally,  the presence of a
corona increases the ratio of the gas pressure to the total pressure
in comparison to the same disc but without a corona. Clearly, the ratio $\beta$ tends to unity at large radial distances. On the other hand, the fraction $f$ of energy exchange is directly proportional to the ratio $\beta$ (see equation (\ref{eq:f})). Thus, the fraction $f$ reaches to its maximum value when the ratio $\beta$ tends to the unity. Thus, the maximum value of $f$ is $\sqrt{2\alpha_{0}}$ according to equation (\ref{eq:f}).

Figure $\ref{fig:figure3}$ shows the typical behavior of  the
physical quantities of the disc for different values of the
viscosity exponent $\mu$. The top left-hand plot of Figure
$\ref{fig:figure3}$ shows that the surface density of the disc
increases when the viscosity exponent increases in both cases with a
corona and without a corona. The presence of a corona enhances the
surface density for all of the input parameters of the viscosity
exponent. However, in the case of a system without a corona, the
profiles are more sensitive to the variations of the viscosity
exponent $\mu$. The behavior of the temperature is shown in the top
right-hand plot of Figure $\ref{fig:figure3}$. The temperature of
the disc increases when viscosity exponent increases for both cases
with and without a corona. Generally, the presence of a corona
serves to cool the disc, but as the viscosity exponent increases the
disc becomes warmer in particular in the inner parts. The bottom
left-hand side of Figure $\ref{fig:figure3}$ shows that the ratio of
disc thickness to the radius increases with increasing values of the
parameter $\mu$, though the enhancement is not significant.  The
presence of a corona causes the disc to become thinner in comparison
with the same disc without a corona for all of the input parameters
of the viscosity exponent. The bottom right-hand plot of Figure
$\ref{fig:figure3}$ shows that the corona causes $\beta$ to increase
in comparison to the case without a corona. When the viscosity
exponent increases then the ratio $\beta$ increases for a disc
without a corona but decreases for the case of a disc with a corona.

The measured values of the magnetic strength of the neutron stars
are in the range $10^{8.5}$ G to $10^{12}$ G (Dewey et al 1986;
Frank, King \& Raine 2002). This implies that the magnetic fields of
the neutron stars in the LMXB are at least as strong. It was shown
that the magnetic field of neutron stars probably do not decay below
$10^{9}$ G or if they do the timescale is larger than $10^9$ yr (Van
den Heuvel et al 1986). Typical behaviors of the physical variables
with different strengths of the stellar  magnetic field are shown in
Figure $\ref{fig:figure4}$. The top left-hand plot of Figure
$\ref{fig:figure4}$ shows the surface density of the disc. The
surface density of the disc increases when the magnetic field
increases for both cases with a corona  and without a corona. The
presence of a corona enhances the surface density.  The top
right-hand plot of Figure $\ref{fig:figure4}$ shows  the disc
temperature increases when magnetic field increases for both cases
with a corona and without a corona. Presence of the corona decreases
temperature of the disc for all of the input parameters. Generally,
we see that the corona becomes more effective when the magnetic
field has larger values. The bottom left-hand plot of Figure
$\ref{fig:figure4}$ shows that the ratio of disc thickness to the
radius increases when magnetic field increases. The corona causes
the disc to become thinner in comparison with the same disc without
a corona, also when magnetic field is stronger, the corona becomes
more effective. The bottom right-hand plot of Figure
$\ref{fig:figure4}$ shows the ratio of gas pressure to the total
pressure of the disc. The ratio $\beta$ decreases when the magnetic
field increases for both cases with a corona and without a corona.
The corona causes ratio of gas pressure to the total pressure of the
disc to increase for all of the input parameters.

The effect of the spin period on the physical parameters is
displayed in  Figure $\ref{fig:figure5}$. The profile of the surface
density is shown in the top left-hand plot of this figure. The
surface density of the disc decreases when the  spin period of the
star increases for both cases with a corona and without a corona.
For all of the  input parameters, the presence of a corona enhances
the surfaces density. The right-hand plot of Figure
$\ref{fig:figure5}$ shows  the corona decreases temperature of the
disc in comparison with the same disc without a corona, irrespective
of the spin period. Also, the temperature of the disc decreases when
the spin period increases and  clearly this reduction of temperature
is more evident  when the star is rotating slower. The bottom
left-hand side of Figure $\ref{fig:figure5}$ shows that when spin
period increases then the ratio of disc thickness to the radius
decreases for all cases. In this regards, the corona becomes more
effective when the star is rotating faster. Also the presence of the
corona causes the disc becomes thinner for all of the input
parameters. The bottom right-hand plot of Figure $\ref{fig:figure5}$
shows that the presence of a corona enhances the ratio of gas
pressure to the total pressure of the disc. The ratio of gas
pressure to the total pressure of the disc increases when spin
period increases and even it tends to unity for spin periods longer
than $5$ seconds which implies negligible radiation pressure.
Generally, we see that the corona becomes more effective when the
spin period has smaller values.

\section{Conclusion}
We  obtained  analytic solutions for the thin accretion disc with corona, to
which a magnetic torque due to the central object (e.g. neutron
star) is applied.
We assumed that all angular momentum transport takes place in  the
disc and the mass accretion rate $\dot{M}$ is constant with radius
and time. In our simple model there is only energy exchange between
the corona and the underlying optically thick disc. We showed that the existence of corona affects
structure  of the disc which is under the influence of the  central
magnetic field. Central in our study is the assumed existence of
magnetic loops anchored in the disc and extending into a corona on
both sides of the disc. Such loops are believed to arise by MRI
inside the disc. We showed that the presence of a corona serves to
cool the disc because of the energy transported from the disc to the
corona. This reduction of the temperature causes the disc becomes
thinner. The surface density increases when we consider the corona
in comparison to the same disc without a corona. Our solutions indicate
that the presence of corona increases ratio of the gas pressure to
the total pressure.

Although we considered a generalized prescription for the viscosity, profiles of the physical variables are weakly depending on the viscosity exponent in both cases with a corona and without a corona. But behaviors of the solutions are highly depend on the viscosity coefficient. The effect of corona is more evident as viscosity tends to high values.  Note that we can not directly compare our solutions with Matthews et
al (2005) study. In fact, our viscosity prescription and viscosity
coefficient are not similar to the analysis of Matthews et al (2005).
Moreover, we considered both radiation and gas pressures. Another important factor in our model is the strength of the central magnetic field at the surface of the neutron star. For low level strength of the surface magnetic field, the effect of corona on profiles of the system is negligible. Also, in this case, the disc is in gas pressure dominated regime. Corona significantly modifies physical profiles of the system with increasing the strength of the central magnetic field. The spin period of the central neutron star is also another important factor in our model. As the star rotates faster, the effect of corona on the behaviors of the physical variables becomes more significant. These findings clearly show the importance of corona in LMXBs which can not be neglected in modeling of such systems. We note that the results are obtained based on only energy exchange from the disc to the corona. Further improvements of our model will include a more accurate description of disc-corona interactions, such as mass exchange or even angular momentum transport. Also, it will be interesting to calculate modifications to the disc spectrum due to the existence of corona according to our analytical solutions.

\acknowledgments
We are grateful to  the anonymous referee for very useful suggestions and comments to improve the paper.
F. K. is grateful for Ad Astra PhD Scholarship of University College of Dublin.
 The research of M. S. was funded under the Programme for Research in
Third Level Institutions (PRTLI) administered by the Irish Higher Education Authority
under the National Development Plan and with partial
support from the European Regional Development Fund.

\end{document}